\documentclass[epj]{svjour}
\usepackage{multirow}

\usepackage{graphics}
\usepackage[utf8]{inputenc}
\usepackage{graphicx}
\usepackage{psfrag}
\usepackage{amssymb}
\usepackage{amsmath}
\usepackage{epstopdf}
\usepackage{color}
\usepackage{booktabs}
\usepackage{bm}
\begin{document}
\title{Revealing the inner structure of the newly observed $\eta_1(1855)$ via photoproduction}
\author{Yin Huang\inst{1}
\and Hong Qiang Zhu\inst{2} \thanks{\emph{Email address:}  20132013@cqnu.edu.cn}
%
}                     
%
%
\institute{School of Physical Science and Technology, Southwest Jiaotong University, Chengdu 610031,China
\and College of Physics and Electronic Engineering, Chongqing Normal Universit}
\date{Received: date / Revised version: date}
%
\abstract{
Very recently, a new hadronic exotic state $\eta_1(1855)$ at the invariant mass spectrum of $\eta\eta^{'}$ was observed by the BESIII Collaboration.
According to its properties, such as the spectroscopy and decay width, the $\eta_1(1855)$ have been suggested to be a compact multi-quark state, a hadron
molecule, or a hybrid meson.  In order to distinguish the various interpretations of the $\eta_1(1855)$, a Reggeized model combined with the vector
dominance model for $\eta_1(1855)$ photoproduction on the proton target is presented.  If the $\eta_1(1855)$ is a $s\bar{s}g$ hybrid meson, the $\eta_1(1855)$
can be produced  though the Primakoff effect.  However, the $\eta_1(1855)$ photoproduction is dominated by the $t$-channel vector mesons $\rho$, $\omega$,
and $\phi$ exchange by assuming $\eta_1(1855)$ as an $S$-wave $K\bar{K}_1(1400)$ molecular state.  Our calculations show that the total cross section of
the $\eta_1(1855)$ production via $\gamma{}p$ reaction can reach up to 0.115 pb, about 0.0124$\%$ of the total cross section obtained by considering the
$\eta_1(1855)$ as an $S$-wave $K\bar{K}_1(1400)$ molecule.  We also find their line shapes are sizably different.  If the $\eta_1(1855)$ is a molecular state,
the photoproduction of $\eta_1(1855)$ near the threshold offers a nice place to test its molecular nature.  However, it should be better to take high energy,
at least above $E_{\gamma}=16.97$ GeV, to observe the production of $\eta_1(1855)$ if $\eta_1(1855)$ is a $s\bar{s}g$ hybrid meson.  These results can be measured
in the GlueX experiment or Electron-Ion Collider in China to test the nature of the $\eta_1(1855)$.
} 
\maketitle

\section{Introduction}\label{sec:intro}
Over the past several decades, many hadrons with different quark components have been observed in experiments~\cite{Zyla:2020zbs}.
These findings have aroused widespread attention, and some theoretical studies on the inner structure have been performed
based on various methods and models.  However, our knowledge of these hadrons is still limited.  Understanding the nature of these
states from different theoretical sides has been and still is a central goal of particle physics.

From the point of the quark model, some of them can be interpreted as mesons made of quark-anti-quark pairs and
baryons constructed from three quarks.  However, there is a plethora of states whose structure is puzzling, presenting properties
that can be fitted in the hadronic molecular state picture if the mass is close to the threshold of two hadrons.  For example,
the first well-known hadronic molecular state is deuteron, composed of neutron-proton component.  Three narrow hidden-charm
pentaquark states, named $P_c(4312)$, $P_c(4440)$, and $P_c(4457)$, have been reported by the LHCb Collaboration~\cite{LHCb:2019kea}.
Their spectroscopies and decay widths can be well explained in the context of the $\Sigma_c\bar{D}^{(*)}$ molecular states~\cite{Chen:2019bip,Guo:2019fdo,Xiao:2019aya,He:2019ify,Xiao:2019mvs}.

Apart from the molecular state, the compact multi-quark and hybrid mesons are the other popular interpretations of the exotic states.
Of particular interest is the hybrid meson that is predicted by Quantum chromodynamics (QCD).  This is because a light hybrid meson may
be composed of one quark and one antiquark together with a few gluons, which leads to the hybrid mesons being a nice place to understand
the interaction between the gluon and the quark.  However, there are only two experimental candidates for light hybrid mesons, named $\pi_{1}(1400)$
and $\pi_1(1600)$, have been recorded in the latest particle data group~\cite{Zyla:2020zbs}. Thus, more studies on the hybrid mesons
are needed.

Fortunately, a meson named $\eta_1(1855)$ was newly observed by the BESIII Collaboration in the analysis of the
$J/\psi\to\gamma\eta\eta^{'}$ reaction~\cite{BESIII:2022riz,BESIII:2022iwi}.  Its mass, width, and favorable quantum numbers were measured
to be
\begin{align}
M&=1855\pm{}9^{+6}_{-1}~~~~~{\rm MeV},\nonumber\\
\Gamma&=188\pm18^{+3}_{-8}~~~~~ {\rm MeV},~~J^{PC}=1^{-+},
\end{align}
respectively.  The exotic quantum number $J^{PC}=1^{-+}$ and the mass $M=1855\pm{}9^{+6}_{-1}$ MeV that the $\eta_1(1855)$ possesses proved
strong evidence for $\eta_1(1855)$ as a good hybrid state candidate.   This is because the existence of a hybrid meson with $J^{PC}=1^{-+}$
and the mass in the 1.8 to 2.1 GeV mass region has been predicted in many early works~\cite{Lacock:1996ny,Meyer:2015eta,Lacock:1996vy,MILC:1997usn,Page:1998gz}.
Indeed,  many works at present can nature interpret the $\eta_1(1855)$ as a hybrid meson.  In particular, the QCD axial anomaly supports
the interpretation of the $\eta_1(1855)$ as the $\bar{s}sg$ hybrid meson~\cite{Chen:2022qpd}, which was also supported by the Lattice
QCD~\cite{Chen:2022isv} and the flux tube model picture~\cite{Qiu:2022ktc}.  An additional hybrid meson explanation of the $\eta_1(1855)$ can be
found in Ref.~\cite{Shastry:2022mhk}.

It is also confusing to see that the observed $\eta_1(1855)$ could be embedded into the compact tetraquark configuration~\cite{Wan:2022xkx,Su:2022eun}.
One reason for this is that the gluon of the hybrid can easily split into a quark pair, which forms a compact tetraquark state.
This imply that it is difficult to differentiate the hybrid mesons from the compact tetraquark states through the strong decay models.
The radiative decay may be a useful way to reveal their difference because the photon cannot interact directly with the gluon.  Based on
the quark model, in which the decay of a hadron is assumed to proceed through a single quark transition, and the total decay width can be
obtained simply by summing the partial widths~\cite{Koniuk:1979vy}, we predict that if the $\eta_1(1855)$ is a compact tetraquark state,
the radiative decay width is two times larger than the results that are obtained by assuming $\eta_1(1855)$ as a hybrid meson$^{[1]}$.
\footnotetext[1]{Here we consider the lowest order Feynman diagram.  That means the $\eta_1(1855)$ only consists of valence quark if $\eta_1(1855)$
is a compact tetraquark state or valence quark and valence gluon if $\eta_1(1855)$ is a hybrid meson.}
Moreover, $\eta_1(1855)$ can also be considered as $K\bar{K}_1(1400)$ bound state~\cite{Yang:2022lwq,Dong:2022cuw}.  Its molecular component
$K\bar{K}_1(1400)$ can be accurately estimated through the radiative decay width~\cite{Yang:2022lwq}.  A possible explanation for this may be
that $\eta_1(1855)$ interacts with the photons through its molecular component $K\bar{K}_1(1400)$, which is quite different from the quark
interacts directly with photons if $\eta_1(1855)$ is a compact tetraquark state or hybrid state.

As discussed above, radiative decay can be employed to judge the various interpretation of the inner structure of the $\eta_1(1855)$.
However, photoproduction of $\eta_1(1855)$ on the nucleon, $\gamma{}p\to{}\eta_1(1855)p$, provides an alternative better way to
test the nature of the $\eta_1(1855)$.  Besides the differences mentioned above in interactions between the photon and molecular
state, hybrid state, and compact tetraquark state, the most important reason is the current experimental information on the
photoproduction of the light meson with masses around 2.0 GeV is abundant.  Thus, searching for the $\eta_1(1855)$ in the
$\gamma{}p\to{}\eta_1(1855)p$ reaction is available.   In this work, we study the photoproduction of the newly observed $\eta_1(1855)$
by assuming $\eta_1(1855)$ as molecular state and hybrid state, respectively.  A definite conclusion on the inner structure of $\eta_1(1855)$
can be  obtained if the obtained cross section could compare with future experimental data.  If the $\eta_1(1855)$ is a hybrid meson state, 
it can be easily distinguished from its compact tetraquark state structure by radiative decay.  Thus, we will not estimate the 
photoproduction cross section by considering the $\eta_1(1855)$ as a compact tetraquark state.

This work is organized as follows. The theoretical formalism is explained in Sec.~\ref{Sec: formulism}. The predicted cross sections are presented
in Sec.~\ref{Sec: results}, followed by a short summary in the last section.

\section{FORMALISM AND INGREDIENTS}\label{Sec: formulism}
In this work, we study the $\gamma{}p\to{}\eta_1(1855)p$ reaction in the framework of an effective Lagrangian approach, which has been
widely employed to investigate the photoproduction processes.  In particular, it has been applied to $\pi$, $\omega$, $\rho$, and $K^{(*)}$
photoproduction at high energies with success.  In the following sections, we first briefly introduce the photoproduction of $\eta_1(1855)$
in the $\gamma{}p\to{}\eta_1(1855)p$ reaction with assuming $\eta_1(1855)$ as $K\bar{K}_1(1400)$ molecule.

\subsection{$\eta_1(1855)$ production as $K\bar{K}_1(1400)$ molecule}
\begin{figure}[h!]
\begin{center}
\includegraphics[bb=60 580 1050 705, clip, scale=0.55]{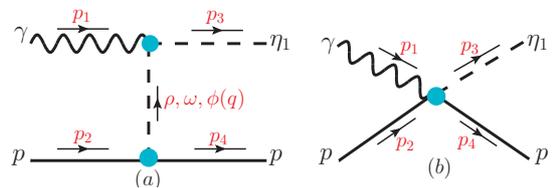}
\caption{Feynman diagrams for the process $\gamma{}p\to{}\eta_1(1855)p$.  The contributions from the $t$-channel vector meson $\rho,\omega,\phi$
exchanges (a) and contact term (b) are considered. We also show the definition of the kinematics ($p_1,p_2,p_3,p_4$) that we use in the present
calculation.}\label{cc1}
\end{center}
\end{figure}
Here, we explain how to calculate the reaction $\gamma{}p\to{}\eta_1(1855)p$ under the $K\bar{K}_1(1400)$ molecular assignment for
$\eta_1(1855)$.  The considered Feynman diagrams are illustrated in Fig.~\ref{cc1}, which includes the $t$-channel vector-meson $\rho$
,$\omega$, $\phi$ exchanges, and the contact diagram.   Note that the contributions from the $s$-channel and the $u$-channel nucleon or
its excited states are ignored because the intermediate nucleon or its excited states will be highly off-shell.  In other words, the
$\gamma{}p\to{}\eta_1(1855)p$ reaction is at high energies beyond the nucleon resonance region, which leads to the contributions
from the $s$-channel and $u$-channel being quite small.  More important is these two channels involve the coupling strength of $\eta_1(1855)$
with the nucleon (or its excited states) is unknown.

To evaluate the diagram shown in Fig.~\ref{cc1}(a), the following effective Lagrangians are needed~\cite{Amundson:1992yp,Oh:2002rb,Sato:1996gk,Rijken:1998yy}$^{[2]}$
\footnotetext[2]{Notice that in our work, the Lagrangian in Eq.~\ref{eq2} is different from Re.~\cite{Amundson:1992yp} where the constants are collected in $g_{\gamma{}V\eta_1}$.}
\begin{align}
{\cal{L}}_{\gamma{}V\eta_1}&=g_{\gamma{}V\eta_1}{\cal{F}}_{\mu\nu}V^{\mu}\eta_1^{\nu}\label{eq2},\\
{\cal{L}}_{VNN}&=-g_{VNN}\bar{N}[\gamma_{\mu}-\frac{\kappa_V}{2m_N}\sigma_{\mu\nu}\partial^{\nu}]V^{\mu}N,
\end{align}
where $\sigma_{\mu\nu}=\frac{i}{2}(\gamma_{\mu}\gamma_{\nu}-\gamma_{\nu}\gamma_{\mu})$, ${\cal{F}}_{\mu\nu}=\partial_{\mu}{\cal{A}}_{\nu}-\partial_{\nu}{\cal{A}}_{\mu}$,
and $\epsilon_{\mu\nu\alpha\beta}$ is the Levi-Civit\`{a} tensor with $\epsilon_{0123}=1$.  $N$, $V$, ${\cal{A}}$, and $\eta_1$ are the nucleon, vector mesons, photon, and $\eta_1(1855)$ meson fields, respectively.   $m_N$ represents the
masses of the nucleon.  The coupling constants for vertexes $\rho{}NN$ and $\omega{}NN$ are taken from the analyses of $\pi$ photoproduction, $\omega$ photoproduction,
nucleon-nucleon scattering, and $\pi-N$ scattering~\cite{Oh:2002rb,Sato:1996gk,Rijken:1998yy}
\begin{align}
&g_{\omega{}NN}=10.35,~~~~~  \kappa_{\omega}=0, \nonumber\\
&g_{\rho{}NN}=6.12,   ~~~~~~~  \kappa_{\rho}=3.1.\nonumber
\end{align}
The $g_{\phi{}NN}$ coupling constant is determined by using the Nijmegen potential as~\cite{Stoks:1999bz,Rijken:1998yy}
\begin{align}
g_{\phi{}NN}=-1.47, ~~~~~\kappa_{\phi}=-1.65,\nonumber
\end{align}
which can describe the hyperon-nucleon scattering data quite well~\cite{Stoks:1999bz,Rijken:1998yy}.  Moreover, the coupling constant $g_{\gamma\eta_1V}$ can be estimated
through the partial decay width of $\Gamma_{\eta_1\to\gamma{}V}$.  At present, there are no clear experimental evidence for the partial decay width of $\Gamma_{\eta_1\to\gamma{}V}$, therefore it should better be determined by theoretical calculation.   By assuming $\eta_1(1855)$ as $K\bar{K}_1$
molecule, the partial widths are evaluated as~\cite{Yang:2022lwq}
\begin{align}
&\Gamma_{\eta_1(1855)\to\gamma{}\rho}=12.63^{+0.14}_{-0.34}~~~~~{\rm KeV},\nonumber\\
&\Gamma_{\eta_1(1855)\to\gamma{}\omega}=12.50^{+0.14}_{-0.34}~~~~~{\rm KeV},\nonumber\\
&\Gamma_{\eta_1(1855)\to\gamma{}\phi}=17.67^{+0.38}_{-0.62}~~~~~{\rm KeV}.\nonumber
\end{align}
Using the Eq.~\ref{eq2}, the two body decay width $\Gamma_{\eta_1\to\gamma{}V}$ is related to $g_{\gamma\eta_1V}$ as
\begin{align}
\Gamma_{\eta_1\to\gamma{}V}=\frac{(m^2_{\eta_1}+m_V^2)(m^2_{\eta_1}-m_V^2)^3}{96\pi{}m_V^2m^5_{\eta_1}}g^2_{\gamma{}V\eta_1}\label{eq4},
\end{align}
which leads to
\begin{align}
g_{\gamma{}\rho\eta_1}=0.0233^{+0.0001}_{-0.0003} ,\nonumber\\
g_{\gamma{}\omega\eta_1}=0.0235^{+0.0001}_{-0.0003},\nonumber\\
g_{\gamma{}\phi\eta_1}=0.0443^{+0.0004}_{-0.0008},\nonumber
\end{align}
where $m_{\eta_1}$ and $m_V$ are the masses of $\eta_{1}(1855)$ and vector meson $V$ ($V=\rho, \omega$, $\phi$), respectively.

Using the effective Lagrangians mentioned above, the invariant amplitude of the $\gamma{}p\to\eta_1(1855)p$ via $t$-channel vector mesons
exchange can be derived as
\begin{align}
{\cal{M}}^{molecule}_{a}&=\sum_{V=\rho,\omega,\phi}{\cal{M}}_V^{\sigma\eta}\epsilon_{\sigma}(p_1,s_1)\epsilon^{\dagger}_{\eta}(p_3,s_3),
\end{align}
with
\begin{align}
{\cal{M}}_V^{\sigma\eta}&=-g_{VNN}g_{\gamma{}V\eta_1}\bar{u}(p_4,s_4)[\gamma_{\mu}+\frac{\kappa_V}{4m_N}(\gamma_{\mu}q\!\!\!/-q\!\!\!/\gamma_{\mu})]u(p_2,s_2)\nonumber\\
             &\times{}\frac{-g^{\mu\alpha}+q^{\mu}q^{\alpha}/m_V^2}{q^2-m_V^2}(p_{1\alpha}g^{\eta\sigma}-p_{1\eta}g^{\alpha\sigma}){\cal{F}}_V(q)\label{eq6},
\end{align}
where $u$ is the Dirac spinor of nucleon, and $\epsilon_{\sigma}$ and $\epsilon^{\dagger}_{\eta}$ are the polarization vector of
photon and $\eta_1(1855)$ meson, respectively.  $s_i(i=1,2,3,4)$ correspond to the spin projection of the  photon, initial proton, $\eta_1(1855)$, and
outgoing proton.  $q$ and $M_V$ are the four-momentum and the masses of the exchanged mesons $\rho$,$\omega$, and $\phi$,
respectively.  ${\cal{F}}_V(q)$ is the form factor.  For the $t$-channel $\rho$, $\omega$, and $\phi$ mesons exchange diagram, we take the
form factor as
\begin{align}
{\cal{F}}_V(q)=\frac{\Lambda_V^2-m_V^2}{\Lambda_V^2-q^2}\label{eq7},
\end{align}
where $\Lambda_{\rho}=\Lambda_{\omega}=1.04$ GeV and $\Lambda_{\phi}=1.8$ GeV, which are chosen to reproduce the vector or pseudoscalar mesons photoproduction~\cite{Kochelev:2009xz}.

At high energies and small momentum transfer, the exchange mechanism for meson photoproduction is dominated by the Regge trajectories in the $t$-channel.
In other words, the usual meson Feynman propagator should be replaced by the Regge propagator.  The adoption of Regge propagators for vector mesons $\rho$,
$\omega$, and $\phi$ exchanges are obtained from Ref.~\cite{Kashevarov:2017vyl} and can be written in the following form
\begin{align}
\frac{1}{t-m_V^2}\Rightarrow{}&=(\frac{s}{s_0})^{\alpha_V(t)-1}\frac{\pi\alpha^{'}}{\sin{[\pi\alpha_V(t)]}}\nonumber\\
                              &\times\frac{{\cal{S}}+e^{-i\pi\alpha_V(t)}}{2}\frac{1}{\Gamma[\alpha_V(t)]},
\end{align}
where $s=(p_1+p_2)^2$ and $t=q^2$ are the Mandelstam variable.  $s_0=1.0$ GeV$^2$ is a mass scale factor.  The signature ${\cal{S}}$ is determined as ${\cal{S}}=(-1)^{J}$
for bosons and  ${\cal{S}}=(-1)^{J+1/2}$ for fermions.  $\Gamma[\alpha(t)]$ is the Gamma function, which is introduced to suppress additional poles of the propagator.
The Regge trajectories are of the form $\alpha_V(t)=\alpha_0+\alpha^{'}t$, where the numerical values of $\alpha_0$ and $\alpha^{'}$ are
taken from Ref.~\cite{Kashevarov:2017vyl}, and given in Table.~\ref{tab11}
\begin{table}[h!]
\begin{center}
\caption{The Regge trajectories used in the present work.}\label{tab11}
\begin{tabular}{cccc} 		 	
  \hline
  \textbf{Regge} ~~~~~~~~&~~~~~ $\alpha_0$      ~~~~~~~~&~~~~~~~~ $\alpha^{'}$(GeV$^{-2}$)~~ \\\hline		
  $\rho$         ~~~~~~~~&~~~~~ 0.477           ~~~~~~~~&~~~~~~~~ $0.885$     ~~ \\
  $\omega$       ~~~~~~~~&~~~~~ 0.434           ~~~~~~~~&~~~~~~~~ $0.923$     ~~ \\
  $\phi$         ~~~~~~~~&~~~~~ 0.421           ~~~~~~~~&~~~~~~~~ $0.557$     ~~ \\
  \hline
  \end{tabular}
 \end{center}
\end{table}

To compute the amplitude of the contact diagram shown in Fig.~\ref{cc1}(b), we need to check the gauge invariance of the photoproduction amplitude.
Technically, the amplitudes that we obtained currently should satisfy the generalized Ward-Takahashi identity
\begin{align}
p_{1\sigma}{\cal{M}}_V^{\sigma}\equiv{}0.
\end{align}
Obviously, the amplitude given in Eq.~\ref{eq6} indeed keeps the full amplitude gauge invariant. Thus, for the present calculation, we
adopt the simplest form for the amplitude of the contact diagram
\begin{align}
{\cal{M}}^{molecule}_b=0.
\end{align}

\subsection{$\eta_1(1855)$ production as hybrid meson}
\begin{figure}[h!]
\begin{center}
\includegraphics[bb=120 595 1050 710, clip, scale=0.75]{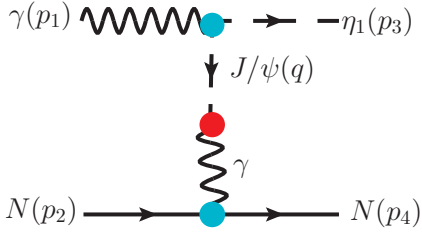}
\caption{Feynman diagram for the process $\gamma{}p\to{}\eta_1(1855)p$ by assuming $\eta_1(1855)$ as a hybrid meson.
The contribution comes from the $t$-channel $J/\psi$ exchange}\label{cc3}
\end{center}
\end{figure}
Considering $\eta_1(1855)$ as a bound state of $s\bar{s}$ and a gluon, the cross section of the reaction $\gamma{}p\to{}\eta_1(1855)p$
can be computed.   Here the photon interacts with the proton through a virtual photon.  Then the virtual photon becomes a $J/\psi$ meson
through the vector meson dominance (VMD) mechanism, which can interact with a photon to produce a $\eta_1(1855)$.  The relevant Feynman diagram is plotted in
Fig.~\ref{cc3}.  To calculate the amplitude of Fig.~\ref{cc3}, it is essential to know the Lagrangians for vertexes $\gamma{}NN$ and
$J/\psi\gamma$.  The Lagrangian for the coupling between photon and nucleon is depicted by
\begin{align}
{\cal{L}}_{\gamma{}pp}=-e\bar{N}[\gamma^{\mu}{\cal{A}}_{\mu}F_1(k^2)-\frac{\kappa_p}{2m_N}\sigma_{\mu\nu}\partial^{\nu}{\cal{A}}^{\mu}F_2(k^2)]N\label{eq11},
\end{align}
where $e=\sqrt{4\pi/137}$ is the unit charge, and the anomalous magnetic momentum of the proton is $\kappa_p=$1.97.
$F_{1,2}(k^2)$ is the proton form factor related to electromagnetic form factors
\begin{align}
G_E(k^2)&=F_1(k^2)+\frac{k^2}{4m_N^2}F_2(k^2),\nonumber\\
G_M(k^2)&=F_1(k^2)+F_2(k^2).
\end{align}
The experimental electromagnetic form factors of the proton can be approximately described by a dipole fit~\cite{He:2005hw}
\begin{align}
G_E(Q^2)=G_M(Q^2)/\mu_p=(1+Q^2/m_D^2)^{-2},
\end{align}
where $m_D^2=0.71$, $Q^2=-k^2$, and $\mu_p=2.793$.

In the VMD mechanism, the Lagrangian depicting the coupling of the intermediate state $J/\psi$ with the photon is written
as
\begin{align}
{\cal{L}}_{J/\psi\gamma}=-\frac{em^2_{J/\psi}}{f_{J/\psi}}J/\psi_{\mu}{\cal{A}}^{\mu}\label{eq14},
\end{align}
where $J/\psi$ and $m_{J/\psi}$ denote the field and the mass of the $J/\psi$ meson, respectively.  $f_{J/\psi}$ is
the decay constant of the vector meson $J/\psi$.  Using the experimental decay width ($\Gamma_{J/\psi\to{}e^{+}e^{-}}=5.529$ KeV)
and the masses of the $m_{J/\psi}=3096.916$ MeV~\cite{Zyla:2020zbs}, one obtains $e/f_{J/\psi}=2.209\times{}10^{-2}$~\cite{Huang:2013mua}.
Moreover, the coupling constant $g_{\gamma{}\eta_1J/\psi}$ can be determined by the decay width of the $J/\psi$
in $\gamma{}\eta_1$ channel.  However, the experimental value of the width $\Gamma_{J/\psi\to\eta_1\gamma}$ at present is unclear.
It can only be determined by the theoretical calculation.  Fortunately, the branching fraction Br($J/\psi\to\gamma\eta_1(1855)$)
is estimated to be about $6.2\times{}10^{-5}$~\cite{Chen:2022isv}.  And the decay width of $J/\psi$ into $\gamma\eta_1(1855)$ is
$\Gamma_{J/\psi\to\gamma\eta_1(1855)}=5.741$ eV  by using the $J/\psi$ total width $\Gamma_{J/\psi}=92.6$ KeV~\cite{Zyla:2020zbs}.
With this value and Eq.~\ref{eq2}, we can obtain
\begin{align}
g_{\gamma{}\eta_1J/\psi}&=\sqrt{\frac{96\pi{}m^2_{\eta_1}m^5_{J/\psi}\Gamma_{J/\psi\to\gamma\eta_1(1855)}}{(m^2_{J/\psi}-m^2_{\eta_1})^3(m^2_{J/\psi}+m^2_{\eta_1})}}\nonumber\\
                        &=7.483\times{}10^{-4}.
\end{align}

Using the interaction Lagrangians shown in Eqs.~\ref{eq2}, \ref{eq11}, and \ref{eq14}, the amplitude related to Fig.~\ref{cc3} can be written as
\begin{align}
{\cal{M}}^{hybrid}&=-i\frac{e^2m^2_{J/\psi}g_{\gamma{}J/\psi\eta_1}}{f_{J/\psi}}\bar{u}(p_4,s_4)[\gamma^{\mu}F_1(q^2)+\frac{\kappa_p}{4m_N}\nonumber\\
                  &\times(\gamma^{\mu}q\!\!\!/-q\!\!\!/\gamma^{\mu})F_2(q^2)]u(p_2,s_2)\frac{-g_{\mu\alpha}+q_{\mu}q_{\alpha}/m^2_{J/\psi}}{q^2(q^2-m^2_{J/\psi})}\nonumber\\
                  &\times{}(p_1^{\alpha}g^{\beta\lambda}-p_{1}^{\beta}g^{\alpha\lambda})\epsilon^{\dagger}_{\beta}(p_3,s_3)\epsilon_{\lambda}(p_1,s_1){\cal{F}}_{J/\psi},
\end{align}
where ${\cal{F}}_{J/\psi}$ is the form factor of the exchanged particle $J/\psi$, and the accurate form will be discussed later.

\section{RESULTS AND DISCUSSIONS}\label{Sec: results}
Based on the theoretical studies in Refs.~\cite{Chen:2022qpd,Chen:2022isv,Qiu:2022ktc,Shastry:2022mhk,Wan:2022xkx,Su:2022eun,Yang:2022lwq,Dong:2022cuw},
$\eta_1(1855)$ can be interpreted as a $\bar{s}sg$ hybrid meson, a compact tetraquark state, and a $K\bar{K}_1$ molecular state, respectively.
Since the photoproduction processes is very helpful to distinguish which inner structure of $\eta_1(1855)$ state is possible, we suggest
experimentally searching for $\eta_1(1855)$ in the $\gamma{}p\to\eta_1(1855)p$ reaction.  With the scattering amplitudes of the
$\gamma{}p\to\eta_1(1855)p$ reaction obtained in the previous section, the differential cross section in the center of mass (c.m.) frame
for the process $\gamma{}p\to\eta_1(1855)p$ can be calculated~\cite{Zyla:2020zbs}
\begin{align}
\frac{d\sigma}{d\cos\theta}=\frac{m_N^2}{32\pi{}s}\frac{|\vec{p}^{ c.m}_3|}{|\vec{p}^{ c.m}_{1}|}\sum_{s_1,s_2,s_3,s_4}|{\cal{M}}^{i}|^2,
\end{align}
where $i$=molecule or hybrid.  The $\theta$ is the scattering angle of the outgoing  $\eta_1(1855)$ meson relative to the beam
direction, while $\vec{p}^{ c.m}_1$ and $\vec{p}^{ c.m}_3$ are the photon and $\eta_1(1855)$ meson three momenta in the C.M. frame, respectively,
which are
\begin{align}
|\vec{p}^{ c.m}_{1}|&=\frac{s-m_p^2}{2\sqrt{s}};\nonumber\\
|\vec{p}^{ c.m}_{3}|&=\frac{\sqrt{[s-(m_{\eta_1}-m_{p})^2][s-(m_{\eta_1}+m_{p})^2]}}{2\sqrt{s}}.
\end{align}

\begin{figure}[h!]
\begin{center}
\includegraphics[bb=-50 100 1050 420, clip, scale=0.45]{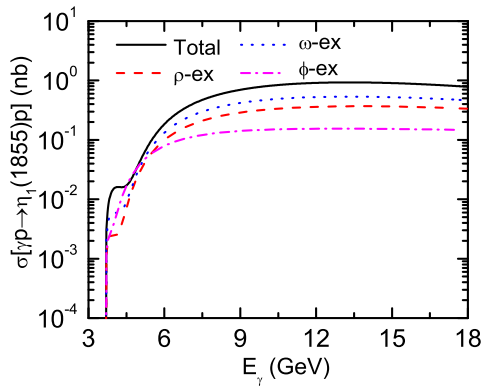}
\caption{The cross section for the $\gamma{}p\to\eta_1(1855)p$ reaction as a function of the beam momentum $E_{\gamma}$
by assuming $\eta_1(1855)$ as a $K\bar{K}_1$ molecular state. With molecular assignment,
the contributions including the $t$-channel $\rho$ exchange (red dash line), $\omega$ exchange (blue dot line), and $\phi$
exchange (magenta dash dot line). The black solid line is the total cross section. }\label{cct1-1}
\end{center}
\end{figure}
We first consider the $\eta_1(1855)$ as a $K\bar{K}_{1}(400)$ molecule.
The cross section $\sigma_{\gamma{}p\to\eta_1(1855)p}[K\bar{K}_1]$ versus the beam momentum of the photon for $\gamma{}p\to\eta_1(1855)p$ reaction
is evaluated.  With molecular state assignment for $\eta_1(1855)$, the production process is described by the $t$-channel $\rho$, $\omega$, and $\phi$ mesons exchange.
Detailed numerical results are shown in Fig.~\ref{cct1-1}.  Here, we only show the cross sections at the center value $g_{\gamma{}\rho\eta_1}=0.0233$, $g_{\gamma{}\omega\eta_1}=0.0235$, and $g_{\gamma{}\phi\eta_1}=0.0443$.

We can find that the total cross section increases sharply near the $\eta_1(1855)p$ threshold.
And such changes can be easily understood due to the phase space opens at that energy.  Then, the total cross section increases continuously but
relatively slowly compared with that near threshold.   To better understand this change, we show the so-obtained total cross section runs from nearly
0.016 nb to 0.929 nb when we change the beam momentum $E_{\gamma}$ from 4.389 GeV to 13.309 GeV.  Theoretically, the cross section can not continuously
increase with $E_{\gamma}$ at high energies as required by unitarity.  Indeed, our results show that the total cross section begins to decrease at a beam
momentum of about $E_{\gamma}=13.309$ GeV.

The individual contributions of $\rho$, $\omega$, and $\phi$ mesons exchange for $\gamma{}p\to\eta_1(1855)p$ reaction as a function of the photon
beam energy are also shown in Fig.~\ref{cct1-1}.   It can be seen that the cross section is the largest for the $\phi$ meson exchange contribution
near the threshold.  With the increase of the beam momentum, the contribution from the $\omega$ meson exchange will plays a predominant role.
At high energy, the following two factors can help readers understand that the contribution from the $\omega$ meson exchange is the most important one.
First, the tensor component related to the $\rho$ meson exchange gives a negative contribution.  Second, although $\eta_1(1855)$ has the largest $\gamma\phi$
decay width, the smallest contribution from the $\phi$ meson exchange can be easily understood due to the $\phi{}NN$ vertex, which involves the creation or
annihilation of an additional $s\bar{s}$ quark pair, is strongly suppressed.  Moreover, the interferences among them are sizable, leading to a bigger total cross section.

Now we turn to the cross section for $\gamma{}p\to\eta_1(1855)p$ reaction by assuming $\eta_1(1855)$ as a $s\bar{s}g$ hybrid meson.  To make a reliable
prediction for the cross section of the $\gamma{}p\to\eta_1(1855)p$ reaction,  the form factor of the exchanged particle $J/\psi$ should be first clarified.
Though the form factor could not be determined from first principles, it can be better determined from the experimental data.  Fortunately, there exists
experimental information on form factor involving $J/\psi$ meson in Ref.~\cite{Wu:2012wta}.  In that work, the experimental data of the process
$\gamma{}p\to{}J/\psi{}p$ can be well reproduced by employing a dipole cutoff function
\begin{align}
{\cal{F}}_{J/\psi}=\frac{\Lambda^2_{J/\psi}}{\Lambda^2_{J/\psi}+\vec{q}^2}~\label{eq19},
\end{align}
with $\Lambda_{J/\psi}=2000$ MeV.  $\vec{q}^2$ is the squared three momentum transfer in the c.m.frame. It is worth noting that the $J/\psi$ in the
$\gamma{}p\to{}J/\psi{}p$ reaction is always on the shell.  And this is different from our work that the $J/\psi$ is considered as an exchanged particle
in the $t$- channel.  But, we still believe that the Eq.~\ref{eq19} can be used to reflect the form factor of $J/\psi$.
\begin{figure}[h!]
\begin{center}
\includegraphics[bb=-50 90 1050 410, clip, scale=0.45]{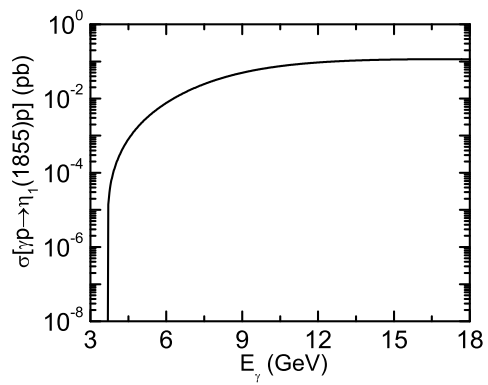}
\caption{The cross section for the $\gamma{}p\to\eta_1(1855)p$ reaction as a function of the beam momentum $E_{\gamma}$
by assuming $\eta_1(1855)$ as a hybrid meson. }\label{cct1}
\end{center}
\end{figure}

By considering the $\eta_1(1855)$ as a $\bar{s}sg$ hybrid meson, the cross section $\sigma_{\gamma{}p\to\eta_1(1855)p}[ssg]$ for the beam
momentum $E_{\gamma}$ from near threshold to 18.0 GeV are shown in Fig.~\ref{cct1}.   We can find that the cross section goes up very
rapidly near the threshold.  However, the value of the cross section increases continuously but relatively slowly at high energy.

Comparing the cross sections shown in Fig.~\ref{cct1-1} and Fig.~\ref{cct1}, we find that if the $\eta_1(1855)$ is a $\bar{s}sg$ hybrid meson,
the cross section is much smaller than the results that are predicted by considering the $\eta_1(1855)$ as a $K\bar{K}_1(1400)$ molecular state.
To see how different the cross sections for the two assignments are, we take the cross section at beam momentum of about $E_{\gamma}=10.0$ GeV
as an example.  Assuming $\eta_1(1855)$ as a $\bar{s}sg$ hybrid meson, the obtained cross section is of the order of 0.0668 pb, while the cross section
can reach 0.801 nb by considering $\eta_1(1855)$ as a $K\bar{K}_1(1400)$ molecule.

Moreover, Fig.~\ref{cct1-1} also tell us that if $\eta_1(1855)$ is a \\$K\bar{K}_1(1400)$ molecular state,  the total cross section of the reaction
$\gamma{}p\to\eta_1(1855)p$ has a clear peak around the $\eta_1(1855)p$ threshold, and the cross section can reach up to 0.0158 nb.  Though the cross
section at this energy is small,  it also indicates that the $\eta_1(1855)$ production near threshold in the $\gamma{}p\to\eta_1(1855)p$ reaction offers
a nice place to test the $K\bar{K}_1(1400)$ molecule interpretations of the $\eta_1(1855)$.  This is because the line shapes of the cross section for the
two assignments of the $\eta_1(1855)$ are sizably different.  Furthermore, we suggest that it will take a high energy, at least above $E_{\gamma}=16.97$ GeV,
to observe the production of $\eta_1(1855)$ in the $\gamma{}p\to\eta_1(1855)p$ reaction if $\eta_1(1855)$ is a $s\bar{s}g$ hybrid meson.  This is because
the cross section of $\gamma{}p\to\eta_1(1855)p$ reaction is the biggest at that energy.

Although the photoproduction cross section for the $\gamma{}p\to\eta_1(1855)p$ reaction that we obtained in this work is quite small, it is still possible
for some experiments to test our theoretical predictions.  In particular, the GlueX experiment at the CEBAF accelerator at Jefferson Lab has accumulated a
larger number of photoproduction data of the light meson with masses around 2 GeV.  Such as a search for a hybrid meson candidate, the $Y(2175)$, in
$\phi(1020)\pi^{+}\pi^{-}$ and $\phi(1020)f_0(980)$ channels in photoproduction on a proton target has been performed~\cite{Hamdi:2020xqv}.  Moreover,
the future electron-ion colliders of high luminosity in China also provide a good platform for searching for $\eta_1(1855)$~\cite{Anderle:2021wcy}.
If the $\eta_1(1855)$ is measured and confirmed in these experiments, a clear conclusion about the nature of the $\eta_1(1855)$ can be verified by comparing
the production cross section predicted in different frameworks.

\section{Summary}\label{sec:summary}

Thanks to the great progress of the experimental, a search for a new hadronic exotic state, the $\eta_1(1855)$, in $\eta\eta^{'}$ channel in the
$J/\psi\to\gamma\eta\eta^{'}$ reaction~\cite{BESIII:2022riz,BESIII:2022iwi} has been performed.  The observed resonance masses, spin-parity, and
widths indicate that the newly observed state $\eta_1(1855)$ is most likely to be a $s\bar{s}g$ hybrid meson.   Indeed, the authors in Refs.~\cite{Chen:2022qpd,Chen:2022isv,Qiu:2022ktc,Shastry:2022mhk} support its interpretations as a $s\bar{s}g$ hybrid meson of $J^{PC}=1^{-+}$.
However, a completely different conclusion was draw from Refs.~\cite{Yang:2022lwq,Dong:2022cuw} that the $\eta_1(1855)$ can be explained as $S$-wave
$K\bar{K}_1(1400)$ molecular state.   It is worth noting that the $\eta_1(1855)$ can be also assigned as compact $s\bar{s}s\bar{s}$ state with $J^{PC}=1^{-+}$~\cite{Wan:2022xkx,Su:2022eun}.  How we sort out the actual component the $\eta_1(1855)$ possess from different theoretical sides is
an urgent question of particle physics.

Theoretical investigations on the production processes will be helpful to distinguish which inner structure of the $\eta_1(1855)$ is possible.
This is because the different production mechanisms of the $\eta_1(1855)$ rely on its structure assignments.  In this work, a Reggeized model combined
with the vector dominance model for $\eta_1(1855)$ photoproduction on the proton target is presented.   The $\eta_1(1855)$ can be produced though the Primakoff effect
if the $\eta_1(1855)$ is a $s\bar{s}g$ hybrid meson.  With assuming $\eta_1(1855)$ as an $S$-wave $K\bar{K}_1(1400)$ molecular state, the
$\eta_1(1855)$ photoproduction is dominated by the $t$-channel vector mesons $\rho$, $\omega$, and $\phi$ exchange.

The numerical results show that the theoretical cross section, which is calculated by assuming $\eta_1(1855)$ as a $s\bar{s}g$ hybrid meson, can reach
0.115 pb.  It is much smaller than the cross section obtained by considering the $\eta_1(1855)$ as an $S$-wave $K\bar{K}_1(1400)$ molecular state.
Moreover, the line shapes of the cross section for the two assignments of the $\eta_1(1855)$ are sizably different.  If the $\eta_1(1855)$ is a molecular
state, the photoproduction of $\eta_1(1855)$ near the threshold is a good way to test its molecular nature.  However, it should be better to take high energy
to observe the production of $\eta_1(1855)$ in the $\gamma{}p\to\eta_1(1855)p$ reaction if $\eta_1(1855)$ is a $s\bar{s}g$ hybrid meson.  These differences
can be easily measured and confirmed in the GlueX experiment or the future electron-ion colliders (EIC) of high luminosity in China.  And is very useful to help
us to test the various interpretations of $\eta_1(1855)$.

\section*{Acknowledgments}
We want to thanks the support from the the National Natural Science Foundation of China under Grant No.12005177,  the Science and Technology Research Program of Chongqing Municipal Education Commission (Grant No. KJQN202200569), the Chongqing Natural Science Foundation project (Grant Nos. cstc2019jcyj-msxmX0560) and the University-level Foundation of Chongqing Normal University (Grant Nos. 21XLB050).

\end{document}